\documentclass[12pt,twoside,a4paper]{posnav}

\usepackage{epsf,exscale,times}
\usepackage[english]{babel}
\usepackage{graphicx}
\usepackage{amsmath}
\usepackage{fancyhdr}
\newcommand {\vsp}   {\vspace*}

\def\title#1{\vsp{-16mm}\begin{center}\Large\bf{#1}\end{center}\vsp{0mm}}
\def\author#1{{\begin{center}\textbf{#1}\end{center}\vspace{-1mm}}}
\def\address#1{\vsp{-3mm}\begin{center}\baselineskip12pt\normalsize{#1}\end{center}\vsp{-1mm}}

\def\abstract#1{{\vspace{-5mm}
    \begin{center}
      \begin{minipage}{0.85\textwidth}
        \noindent\bf \textit{Abstract:}
        \small\rm\emph{#1}
				\vsp{-0.5em}
      \end{minipage}
    \end{center}
}}

\def\authorsheadline=#1{\global\def\@authorsheadline{#1}}
\global\def\@authorsheadline{}

\def\TeX{T\kern-.1667em\lower.5ex\hbox{E}\kern-.125emX}

\def\LaTeXG{{\rm L\kern-.36em\raise.3ex\hbox{\sc a}\kern-.15emT\kern-.1667em\lower.7ex\hbox{E}\kern-.125emX}}
\def\LaTeXK{{\it L\kern-.24em\raise.4ex\hbox{\scriptsize \it A}\kern-.20emT\kern-.1667em\lower.5ex\hbox{E}\kern-.125emX}}

\usepackage{epsfig}
\usepackage[T1]{fontenc}
\usepackage{booktabs}
\usepackage[misc]{ifsym}

\usepackage{paralist}
\usepackage{eurosym}
\usepackage{flushend}
\usepackage{amssymb}
\usepackage{url}
\usepackage[utf8]{inputenc}
\usepackage{amsfonts}
\usepackage{soul}
\usepackage[justification=centering]{caption}
\usepackage[list=true]{subcaption}
\usepackage{tabu}
\usepackage{wrapfig}
\usepackage{arydshln}
\usepackage[normalem]{ulem}
\usepackage{multirow}
\usepackage{siunitx}
\usepackage{hyperref}
\usepackage{makecell}
\usepackage{adjustbox}
\usepackage{array}
\usepackage{diagbox}
\usepackage{dsfont}

\begin{document}

\fancypagestyle{firststyle}
{
   \fancyhf{}
   \lfoot{ \footnotesize{Positioning and Navigation for Intelligent Transport Systems POSNAV 2024, October 1 - 2, 2024, Weimar\break     		\copyright  2024 DGON} }
   \rfoot{ \footnotesize {\thepage} }
}

\thispagestyle{firststyle}
\fancyhf{}
\renewcommand{\headrulewidth}{0pt}
\renewcommand{\footrulewidth}{1pt}
\renewcommand{\footskip}{50pt}

\pagestyle{fancy}
\fancyfoot[RO,LE]{ \footnotesize {\thepage} }

\title{Achieving Generalization in Orchestrating GNSS Interference Monitoring Stations Through Pseudo-Labeling}

\author{
Lucas Heublein, Tobias Feigl, Alexander Rügamer, Felix Ott}
\address{
    Fraunhofer Institute for Integrated Circuits IIS, Nürnberg, Germany\\
	email: \{lucas.heublein, felix.ott\}@iis.fraunhofer.de
	}
\abstract{
The accuracy of global navigation satellite system (GNSS) receivers is significantly compromised by interference from jamming devices. Consequently, the detection of these jammers are crucial to mitigating such interference signals. However, robust classification of interference using machine learning (ML) models is challenging due to the lack of labeled data in real-world environments. In this paper, we propose an ML approach that achieves high generalization in classifying interference through orchestrated monitoring stations deployed along highways. We present a semi-supervised approach coupled with an uncertainty-based voting mechanism by combining Monte Carlo and Deep Ensembles that effectively minimizes the requirement for labeled training samples to less than 5\% of the dataset while improving adaptability across varying environments. Our method demonstrates strong performance when adapted from indoor environments to real-world scenarios. \textbf{Datasets:} \href{https://gitlab.cc-asp.fraunhofer.de/darcy_gnss}{https://gitlab.cc-asp.fraunhofer.de/darcy\_gnss}
}
\section{Introduction}
\label{label_introduction}

\begin{figure*}[!t]
    \centering
    \includegraphics[width=1.0\linewidth]{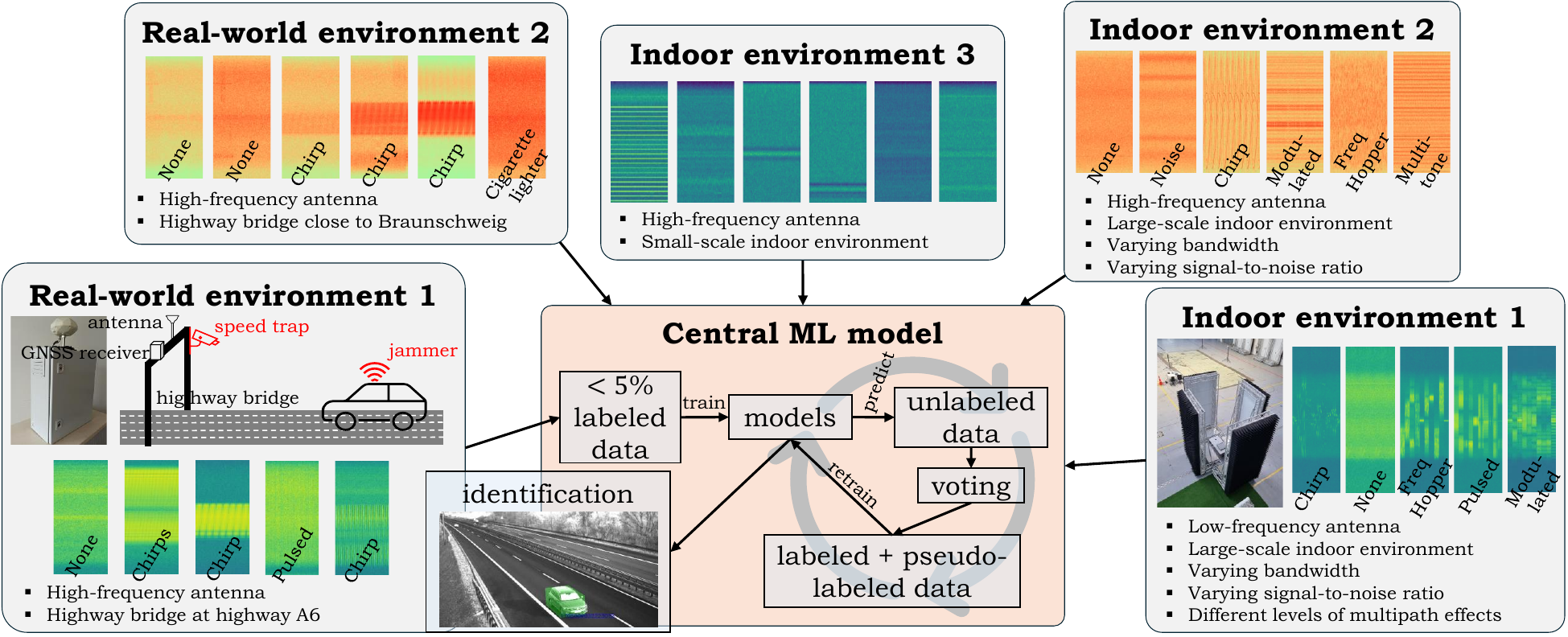}
    \vspace{-0.75cm}
    \caption{An overview of the orchestration of GNSS interference monitoring stations and the reduction of data discrepancies between environments using semi-supervised learning.}
    \label{figure_intro_overview}
\end{figure*}

The localization accuracy of GNSS receivers is significantly degraded by interference signals generated by jamming devices~\cite{crespillo_ruiz}. This issue has become increasingly prevalent in recent years~\cite{ainonline} due to the widespread availability of low-cost and easily accessible jamming devices~\cite{mehr_minetto_dovis,miguel_chen_lo}. As a result, it is imperative to mitigate these interference signals or eliminate the transmitter. This necessitates the detection, classification, and localization of the interference source~\cite{raichur_ion_gnss,raichur_heublein,merwe_franco,brieger_ion_gnss,jdidi_brieger}. One critical application of GNSS interference monitoring is in the management of toll collection for trucks on highways~\cite{ott_heublein_icl}. Most modern systems are electronic; however, GNSS signals can be jammed, hindering the detection of affected trucks. Therefore, it is crucial to detect the jammer within the affected vehicle. To address this, we built a sensor station equipped with a GNSS receiver and antenna on German highways to record GNSS snapshots~\cite{heublein_raichur_ion} (refer to Figure~\ref{figure_intro_overview}) and classify these snapshots to identify the affected vehicle. Our objective is to develop resilient ML models. The construction of GNSS interference monitoring stations with robust ML models presents significant challenges due to the high variability in jamming devices, interference characteristics, sensor and antenna properties, distances between jammers and sensor stations, satellite constellations, and multipath effects from varying environments~\cite{heublein_feigl_crpa}. To enhance ML model accuracy, we employ federated learning to share data, such as interference snapshots, among orchestrated monitoring stations. These models can identify new interference types, share information through the aggregation of model weights, and adapt to and classify these interferences across all sensor stations.

In real-world environments, challenges arise due to missing and unbalanced data, which may result from a lack of interference classes, specific interference characteristics, or labeled data in new environments. Since the use of jamming devices is prohibited, recording specific data in each new environment is not feasible, leading to a scarcity of labeled data. However, it is possible to simulate interferences or record a wide range of interference characteristics in controlled indoor environments (refer to Figure~\ref{figure_intro_overview}). Nonetheless, the domain gap between data collected in indoor settings and real-world outdoor scenarios must be minimized~\cite{heublein_raichur_ion}. One potential solution is semi-supervised learning~\cite{engelen_hoos}, which integrates pre-training using a substantial quantity of labeled data from pre-existing indoor datasets with a smaller set of labeled data from new environments. This approach uses labeled data to guide the learning process while leveraging the abundant unlabeled data to enhance model performance. The core idea is to exploit the structure of the unlabeled data to improve learning, making it particularly valuable in scenarios where labeled data is scarce or costly to obtain~\cite{berthelot_carlini}, such as in GNSS interference monitoring~\cite{jdidi_brieger,wang_chen_guo}. In the technique of pseudo-labeling~\cite{bonilla_tan_yi,park_kim,yan_hui_li,zhu_shi_feng,rizve_duarte}, the model generates labels for the unlabeled data based on its predictions, which are then used to further train the model, leading to improved generalization. However, pseudo-labeling has not yet been applied to GNSS interference monitoring. Figure~\ref{figure_intro_overview} illustrates our proposed pseudo-labeling method. We pre-train our ML model on indoor datasets with high variance in interference characteristics and then adapt it to novel real-world environments (e.g., highways) with only a few labeled samples (i.e., five). Our approach involves pseudo-labeling the unlabeled data and retraining the models using an uncertainty-based voting mechanism by utilizing Monte Carlo~\cite{gal_ghahramani}, Deep Ensembles~\cite{lakshminarayanan}, or a combination of both, resulting in effective and robust adaptation to new environments.

\paragraph{Contributions.} The primary objective of this work is to adapt GNSS interference datasets recorded along highways using only a few samples, leveraging pseudo-labeling while accounting for real-world data discrepancies. We present the following key contributions: (1) We are the first to address the challenge of adapting GNSS interference monitoring from indoor scenarios to real-world environments. (2) We propose a novel pseudo-labeling method that enables adaptation with a minimal number of five positive interference labels. (3) We introduce a voting mechanism by combining Monte Carlo and Deep Ensembles for pseudo-labeling, which uses a softmax threshold to minimize model uncertainty when labeling unlabeled samples. (4) Our evaluation demonstrates that our method outperforms state-of-the-art techniques~\cite{rizve_duarte}, enabling effective adaptation using less than 5\% of labeled data.

%\paragraph{Outlook.} The remainder of this paper is structured as follows: Section~2 offers an overview of the existing literature on GNSS interference monitoring and semi-supervised learning. In Section~3, we describe our pseudo-labeling method, which employs an uncertainty-based voting mechanism. Section~4 presents the experiments conducted and summarizes the evaluation results. Finally, Section~5 provides the concluding remarks.
\section{Related Work}
\label{label_related_work}

Crespillo et al.~\cite{crespillo_ruiz} proposed a logistic regression approach to mitigate biased estimation, thereby enhancing ML generalization. Merwe et al.~\cite{merwe_franco} introduced a low-cost GNSS interference detection and classification receiver. Brieger et al.~\cite{brieger_ion_gnss} considered both spatial and temporal relationships when fusing snapshot and bandwidth-limited data with a joint loss function. Raichur et al.~\cite{raichur_ion_gnss} proposed a crowdsourcing approach with smartphone-based features to localize interference sources. Jdidi et al.~\cite{jdidi_brieger} developed a quasi-unsupervised method that does not rely on prior knowledge and adapts to environment-specific factors such as multipath, dynamics, and signal strength variations. The input data undergoes preprocessing, embedding, disentanglement, and clustering using the K-means algorithm. Subsequently, a random forest classifier is trained utilizing a distance-based label generator. For adapting to new interference types, Ott et al.~\cite{ott_heublein_icl} proposed a few-shot learning approach based on uncertainty and pairwise learning, while Raichur et al.~\cite{raichur_heublein} contributed a Bayesian learning-based dynamic weighting of contrastive and classification loss functions for class-incremental learning. In a previous work~\cite{heublein_feigl_crpa}, we introduced a GNSS dataset featuring a range of interference characteristics, which we utilize for training purposes. Heublein et al.~\cite{heublein_raichur_ion} illustrated the discrepancy in data between indoor and outdoor datasets when evaluating feature embeddings. Wang et al.~\cite{wang_chen_guo} proposed a domain adaptation framework that utilizes labeled GNSS data from a source domain to enhance positioning accuracy in a target domain. Park et al.~\cite{park_kim} introduced a pseudo-labeled method for classifying target contents by incorporating noise labels. Yan et al.~\cite{yan_hui_li} proposed a deep metric learning-based pseudo-labeling technique to develop compact feature representations of both labeled and unlabeled data, addressing class imbalance issues. In the context of fault diagnosis, Zhu et al.~\cite{zhu_shi_feng} utilized a multiple kernel variant of maximum mean discrepancy to assess the discrepancy in marginal probability distributions and pseudo-labels, as well as to evaluate conditional probability distribution discrepancies. They addressed the challenge of pseudo-label noise interference by developing a method to filter out low-quality pseudo-labels using an adaptive threshold. Despite notable progress, challenges persist in comprehensively addressing domain discrepancies in GNSS data.
\section{Methodology}
\label{label_methodology}

\begin{figure*}[!t]
    \centering
    \includegraphics[width=0.7\linewidth]{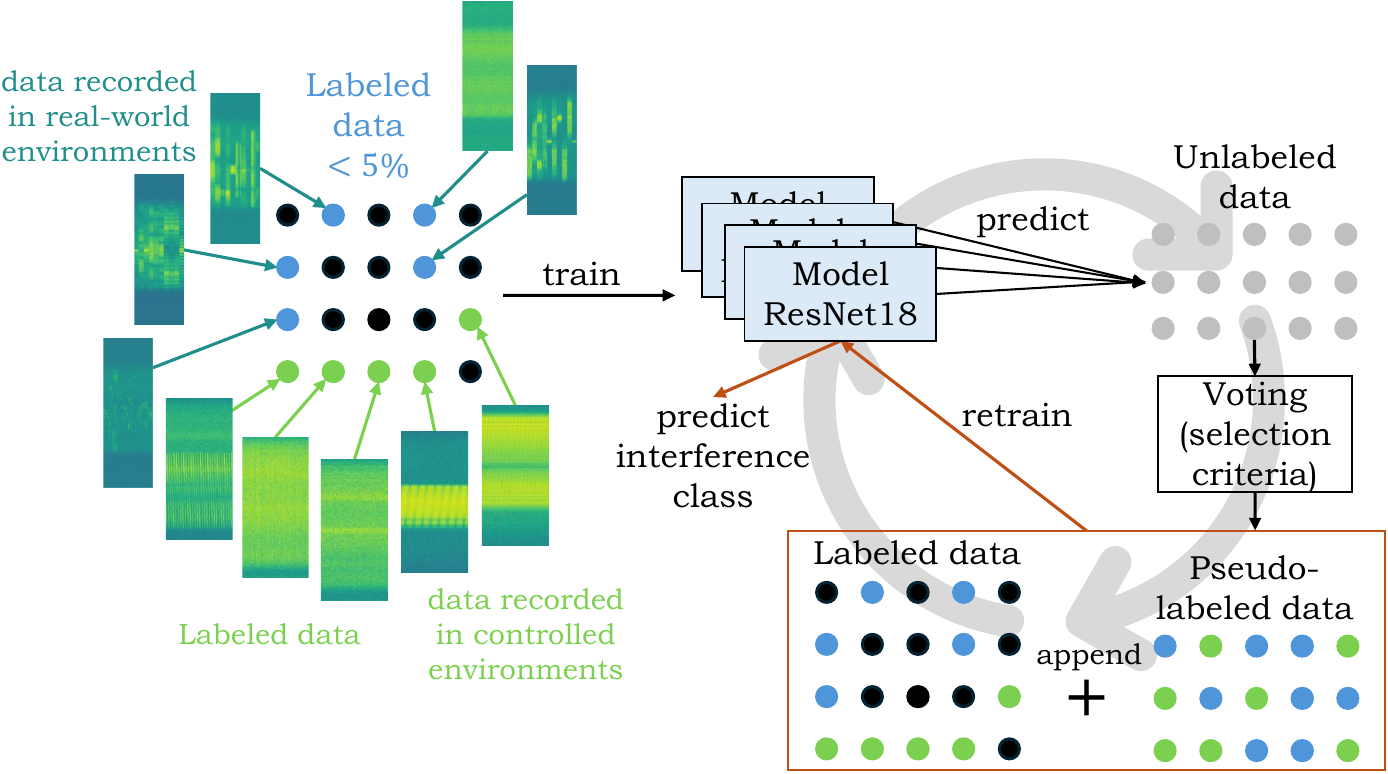}
    \caption{ML pipeline with pseudo-labeling between labeled and unlabeled datasets.}
    \label{figure_pseudo_labeling}
\end{figure*}

\paragraph{Method Overview.} Figure~\ref{figure_pseudo_labeling} presents a comprehensive overview of the proposed method. Initially, a ResNet18~\cite{he_zhang} model, which has demonstrated high performance in GNSS classification tasks~\cite{ott_heublein_icl,heublein_raichur_ion,heublein_feigl_crpa}, is pre-trained on a dataset collected in an indoor environment. The objective is to adapt this model to new scenarios, specifically a highway environment, using only a limited number of labeled samples. We evaluate the model using 0.5\%, 1\%, and 5\% of labeled samples from the original dataset. Subsequently, $M$ ResNet18 models, denoted as $m_i$ where $i \in \{1, 2, \ldots, M\}$, are trained. Experiments are conducted with $M \in \{1, 2, 4\}$ models. The models then predict the class for the remaining unlabeled dataset using a voting mechanism, with selection criteria based on uncertainty estimates from Monte Carlo~\cite{gal_ghahramani} and Deep Ensembles~\cite{lakshminarayanan}, as detailed in the following sections. These pseudo-labeled data are subsequently combined with the labeled dataset, and the models are retrained. Finally, the models perform predictions on the test dataset for the interference classification task.

\paragraph{Pseudo-Label Selection.} Our selection criteria are based on the pseudo-labeling approach proposed by Rizve et al.~\cite{rizve_duarte}, which employs negative learning on hard samples to reduce the noise encountered by the network. The primary goal is to achieve an output distribution that is invariant to perturbations and augmentations. The following notation is used for pseudo-label selection. Let $D_L = \{(\mathbf{X}_i, \mathbf{y}_i)\}_{i=1}^{N_L}$ represent a labeled GNSS dataset with $N_L$ sample snapshots $\mathbf{X}_i$, where $\mathbf{y}_i = [y_i^1, \ldots, y_i^\mathcal{C}] \subseteq \{0,1\}^\mathcal{C}$ denotes the corresponding labels, with $\mathcal{C} = 7$ class categories. Let $D_U = \{\mathbf{X}_i\}_{i=1}^{N_U}$ represent an unlabeled dataset with $N_U$ samples lacking labels, for which pseudo-labels $\hat{\mathbf{y}}_i$ are generated. We train a model $f_\theta$ with network parameters $\theta$ on the combined dataset $\hat{D} = \{(\mathbf{X}_i, \hat{\mathbf{y}}_i)\}_{i=1}^{N_L + N_U}$, with $\hat{\mathbf{y}}_i = \mathbf{y}_i$ for the $N_L$ labeled samples~\cite{rizve_duarte}. Hard pseudo-labels are derived directly from network predictions. Let $\mathbf{p}_i$ denote the probability outputs of a trained network for the sample $\mathbf{X}_i$, where $p_i^c$ represents the probability that class $c$ is present in the sample. The pseudo-label for $\mathbf{X}_i$ can then be generated as $\hat{y}_i^c = \mathds{1}[p_i^c \geq \gamma]$, where $\gamma \in (0,1)$ is a softmax threshold~\cite{rizve_duarte}. The objective is to select a subset of pseudo-labels that are less noisy by choosing those associated with high-confidence predictions. The selected pseudo-labels are denoted as $\mathbf{g}_i = [g_i^1, \ldots, g_i^\mathcal{C}] \subseteq \{0,1\}^\mathcal{C}$, where $g_i^c = 1$ if $\hat{y}_i^c$ is selected and $g_i^c = 0$ if $\hat{y}_i^c$ is not selected. This is obtained using the following equation:
\begin{equation}
    \label{equ1}
    g_i^c = \mathds{1}[p_i^c \leq \tau_p] + \mathds{1}[p_i^c \geq \tau_n],
\end{equation} 
where $\tau_p$ and $\tau_n$ are confidence thresholds for positive and negative labels, respectively~\cite{rizve_duarte}. For our multi-label classification task, we train the parameterized model $f_\theta$ using a binary cross-entropy loss on the selected pseudo-labels.

\paragraph{Monte Carlo Pseudo-Label Selection.} Poor model calibration can result in incorrect predictions that are assigned high confidence scores. Given the direct relationship between the expected calibration error (ECE) score and prediction uncertainties, pseudo-labels associated with predictions that meet certain criteria contribute to a reduced calibration error. Therefore, we employ the Monte Carlo (MC) method, proposed by Gal et al.~\cite{gal_ghahramani} and previously utilized by Rizve et al.~\cite{rizve_duarte}, to select pseudo-labels for which the model exhibits high confidence:
\begin{equation}
    \label{equ2}
    g_i^c = \mathds{1}\big[u(p_i^c) \leq \kappa_p\big] \mathds{1}\big[p_i^c \geq \tau_p\big] + \mathds{1}\big[u(p_i^c) \leq \kappa_n\big] \mathds{1}\big[p_i^c \geq \tau_n\big],
\end{equation} 
where $u(p)$ represents the uncertainty of a prediction $p$. This approach ensures that the network's prediction is sufficiently certain to be selected~\cite{rizve_duarte}. We set the uncertainty thresholds as $\kappa_p = 0.05$ and $\kappa_n = 0.005$, and the confidence thresholds as $\tau_p = 0.70$ and $\tau_n = 0.05$. Notably, the voting is conducted using a single model ($M=1$) with $C$ samples $k_l$, where $l \in \{0, \ldots, C\}$. In the following, we present a method that employs multiple models for voting.

\begin{figure*}[!t]
    \centering
    \includegraphics[width=0.7\linewidth]{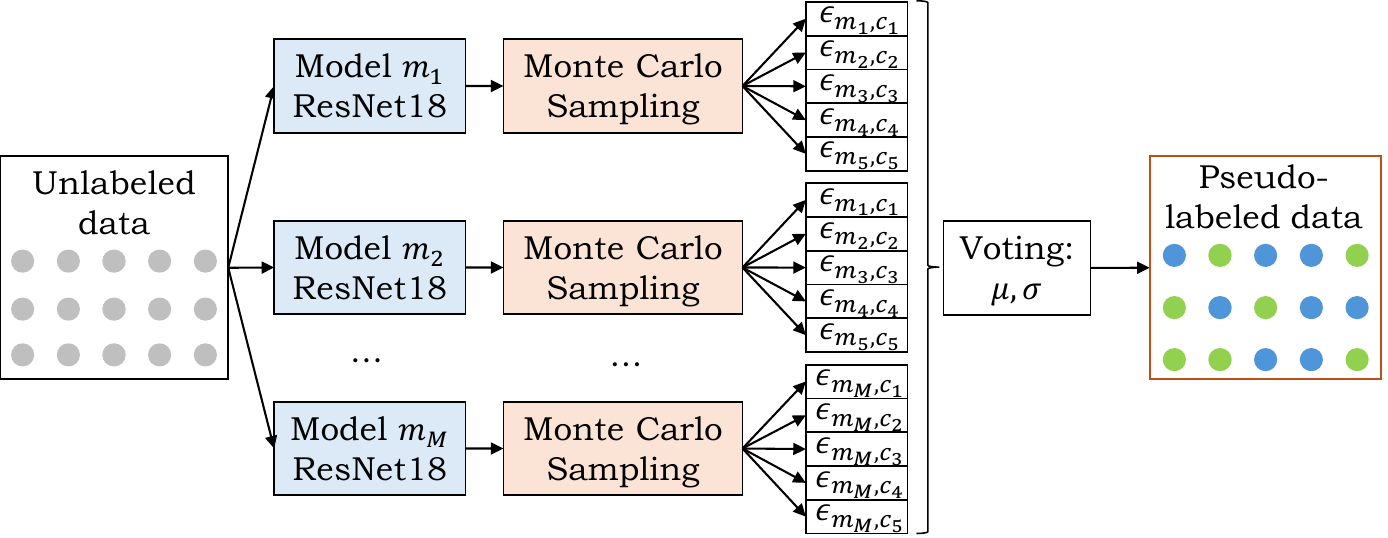}
    \vspace{-0.2cm}
    \caption{The voting process for the unlabeled dataset is conducted based on the mean and standard deviation derived from $M$ models and five Monte Carlo samples.}
    \label{figure_voting}
\end{figure*}

\paragraph{Deep Ensembles Pseudo-Label Selection.} While pseudo-labeling is both versatile and modality-agnostic, it exhibits relatively poor performance compared to recent semi-supervised methods due to the large number of incorrectly pseudo-labeled samples. However, MC dropout requires the model to include a dropout layer. Since our ResNet18 model lacks an integrated dropout layer -- adding which would degrade performance -- we instead employ Deep Ensembles~\cite{lakshminarayanan}, training $M$ identical models without dropout but with different initial weights. Through the $M$ models, we can estimate uncertainty, which involves computing the posterior distribution $p(\theta|D)$ -- the model weights $\theta$ given the training dataset $D$. However, due to the typical intractability of calculating the posterior directly, an approximation is often used. We employ Deep Ensembles, consisting of a committee of $M$ models, each initialized with a unique seed, where the initialization serves as the sole source of stochasticity in the model parameters. The final results are obtained by aggregating predictions from these $M$ models. Our goal is to further reduce the noise in training by integrating the previously introduced MC sampling method with Bayesian inference. Figure~\ref{figure_voting} illustrates the pseudo-labeling process via voting among the $M$ models with MC sampling.

\paragraph{Voting.} If the $M$ models do not agree on the class label corresponding to a given sample, that sample is excluded from use as a pseudo-label in the subsequent retraining phase. Conversely, when all $M$ models predict the same class label, the mean of the softmax outputs $\epsilon_{m_i,k_l}$ for the model $m_i$ and $C$ Monte Carlo samples $k_l$, where $l \in \{0, \ldots, C\}$ of each model $m_i$ must exceed a predefined threshold. We set $C = 5$. The optimal threshold is determined within $[0.7, 0.9, 0.99]$ based on $M \times C$ samples.

\section{Evaluation}
\label{label_evaluation}

As introduced in Figure~\ref{figure_intro_overview}, we transition from indoor environments to outdoor highway scenarios. We begin by providing details on these datasets. Subsequently, we assess the impact of uncertainty (i.e., MC dropout and Deep Ensembles) on our pseudo-labeling approach. Finally, we present an ablation study and conduct a hyperparameter search. All experiments were performed using Nvidia Tesla V100-SXM2 GPUs with 32 GB VRAM, supported by Core Xeon CPUs and 192 GB RAM. We employed the standard SGD optimizer with a learning rate of $0.01$, a decay rate of $5 \times 10^{-4}$, a momentum of $0.9$, and a batch size of 64. The pre-training was conducted over 200 epochs, utilizing a multistep learning rate schedule with milestones at $[120, 160]$ and a gamma of $0.1$. For the pseudo-labeling, training was conducted over 100 epochs, utilizing the milestones at $[60, 80]$ for the initial training and 20 epochs with milestones at $[12, 16]$ for each further epoch. We report the final epoch accuracy in \%.

\begin{table}[b!]
\begin{center}
\setlength{\tabcolsep}{2.4pt}
    \caption{Overview of GNSS datasets and their number of samples for each interference class.}
    \label{table_data_combination}
    \vspace{-0.1cm}
    \footnotesize \begin{tabular}{ p{1.3cm} | p{0.5cm} | p{0.5cm} | p{0.5cm} | p{0.5cm} | p{0.5cm} | p{0.5cm} | p{0.5cm} | p{0.5cm} | p{0.5cm} | p{0.5cm} | p{0.5cm} }
    \multicolumn{1}{c|}{\textbf{Dataset}} & \multicolumn{1}{c|}{\textbf{Antenna}} & \multicolumn{1}{c|}{\textbf{Set}} & \multicolumn{1}{c|}{\textbf{0}} & \multicolumn{1}{c|}{\textbf{1}} & \multicolumn{1}{c|}{\textbf{2}} & \multicolumn{1}{c|}{\textbf{3}} & \multicolumn{1}{c|}{\textbf{4}} & \multicolumn{1}{c|}{\textbf{5}} & \multicolumn{1}{c|}{\textbf{6}} & \multicolumn{1}{c|}{\textbf{7}} & \multicolumn{1}{c}{\textbf{8}} \\ \hline
    \multicolumn{1}{l|}{Controlled large-scale 1+2} & \multicolumn{1}{c|}{low} & \multicolumn{1}{c|}{train} & \multicolumn{1}{r|}{17,764} & \multicolumn{1}{r|}{15,512} & \multicolumn{1}{r|}{47,912} & \multicolumn{1}{r|}{11,320} & \multicolumn{1}{r|}{984} & \multicolumn{1}{r|}{14,268} & \multicolumn{1}{r|}{12,900} & \multicolumn{1}{r|}{} & \multicolumn{1}{r}{} \\
    \multicolumn{1}{l|}{} & \multicolumn{1}{l|}{} & \multicolumn{1}{c|}{test} & \multicolumn{1}{r|}{4,448} & \multicolumn{1}{r|}{3,904} & \multicolumn{1}{r|}{12,004} & \multicolumn{1}{r|}{2,832} & \multicolumn{1}{r|}{248} & \multicolumn{1}{r|}{1,088} & \multicolumn{1}{r|}{3,228} & \multicolumn{1}{r|}{} & \multicolumn{1}{r}{} \\
    \multicolumn{1}{l|}{Controlled large-scale 2} & \multicolumn{1}{c|}{high} & \multicolumn{1}{c|}{train} & \multicolumn{1}{r|}{209,218} & \multicolumn{1}{r|}{106,584} & \multicolumn{1}{r|}{237,524} & \multicolumn{1}{r|}{122,862} & \multicolumn{1}{r|}{10,930} & \multicolumn{1}{r|}{21,266} & \multicolumn{1}{r|}{139,669} & \multicolumn{1}{r|}{} & \multicolumn{1}{r}{} \\
    \multicolumn{1}{l|}{} & \multicolumn{1}{l|}{} & \multicolumn{1}{c|}{test} & \multicolumn{1}{r|}{52,646} & \multicolumn{1}{r|}{27,388} & \multicolumn{1}{r|}{58,583} & \multicolumn{1}{r|}{30,404} & \multicolumn{1}{r|}{2,388} & \multicolumn{1}{r|}{5,476} & \multicolumn{1}{r|}{35,029} & \multicolumn{1}{r|}{} & \multicolumn{1}{r}{} \\
    \multicolumn{1}{l|}{Real-world highway 1} & \multicolumn{1}{c|}{high} & \multicolumn{1}{c|}{train} & \multicolumn{1}{r|}{157,259} & \multicolumn{1}{r|}{7} & \multicolumn{1}{r|}{75} & \multicolumn{1}{r|}{} & \multicolumn{1}{r|}{} & \multicolumn{1}{r|}{} & \multicolumn{1}{r|}{30} & \multicolumn{1}{r|}{78} & \multicolumn{1}{r}{8} \\
    \multicolumn{1}{l|}{} & \multicolumn{1}{l|}{} & \multicolumn{1}{c|}{test} & \multicolumn{1}{r|}{39,315} & \multicolumn{1}{r|}{2} & \multicolumn{1}{r|}{20} & \multicolumn{1}{r|}{} & \multicolumn{1}{r|}{} & \multicolumn{1}{r|}{} & \multicolumn{1}{r|}{11} & \multicolumn{1}{r|}{20} & \multicolumn{1}{r}{2} \\
    \multicolumn{1}{l|}{Real-world highway 2} & \multicolumn{1}{c|}{high} & \multicolumn{1}{c|}{train} & \multicolumn{1}{r|}{12,096} & \multicolumn{1}{r|}{} & \multicolumn{1}{r|}{800} & \multicolumn{1}{r|}{} & \multicolumn{1}{r|}{} & \multicolumn{1}{r|}{} & \multicolumn{1}{r|}{} & \multicolumn{1}{r|}{} & \multicolumn{1}{r}{} \\
    \multicolumn{1}{l|}{} & \multicolumn{1}{l|}{} & \multicolumn{1}{c|}{test} & \multicolumn{1}{r|}{3,024} & \multicolumn{1}{r|}{} & \multicolumn{1}{r|}{200} & \multicolumn{1}{r|}{} & \multicolumn{1}{r|}{} & \multicolumn{1}{r|}{} & \multicolumn{1}{r|}{} & \multicolumn{1}{r|}{} & \multicolumn{1}{r}{} \\
    \end{tabular}
    \vspace{-0.5cm}
\end{center}
\end{table}

\paragraph{Datasets.} We utilize two indoor datasets. The \textit{controlled large-scale dataset 1+2}~\cite{heublein_feigl_crpa}, recorded with a low-frequency antenna, contains GNSS snapshots across six interference classes (see Figure~\ref{figure_intro_overview}, bottom right). Over a period of 24 days, we continuously recorded snapshots with interferences exhibiting varying bandwidth, signal-to-noise ratio, and multipath effects in different scenarios. Concurrently, we recorded the \textit{controlled large-scale dataset 2}~\cite{heublein_raichur_ion} using a high-frequency antenna (see Figure~\ref{figure_intro_overview}, top right). The \textit{small-scale dataset}~\cite{brieger_ion_gnss} (see Figure~\ref{figure_intro_overview}, top middle) was not utilized in our experiments. We adapt to two separate highway datasets, where interference classes were recorded using handheld jammers and jammers in cigarette lighters. The \textit{highway dataset 1}~\cite{ott_heublein_icl} was recorded along the A6 in Germany using a high-frequency antenna (see Figure~\ref{figure_intro_overview}, left). The \textit{highway dataset 2}~\cite{heublein_raichur_ion} was recorded near Braunschweig, also with a high-frequency antenna (see Figure~\ref{figure_intro_overview}, top left). Table~\ref{table_data_combination} provides an overview of the number of samples per interference class and highlights the class imbalance. Due to the significant overlap between class labels, we restrict our analysis to only the classes 0 (non-interference) and 1 (\textit{chirp}).

\newcommand\y{0.325}
\begin{figure*}[!t]
    \centering
	\begin{minipage}[t]{1.0\linewidth}
    	\begin{minipage}[t]{\y\linewidth}
            \centering
        	\includegraphics[trim=10 10 10 10, clip, width=1.0\linewidth]{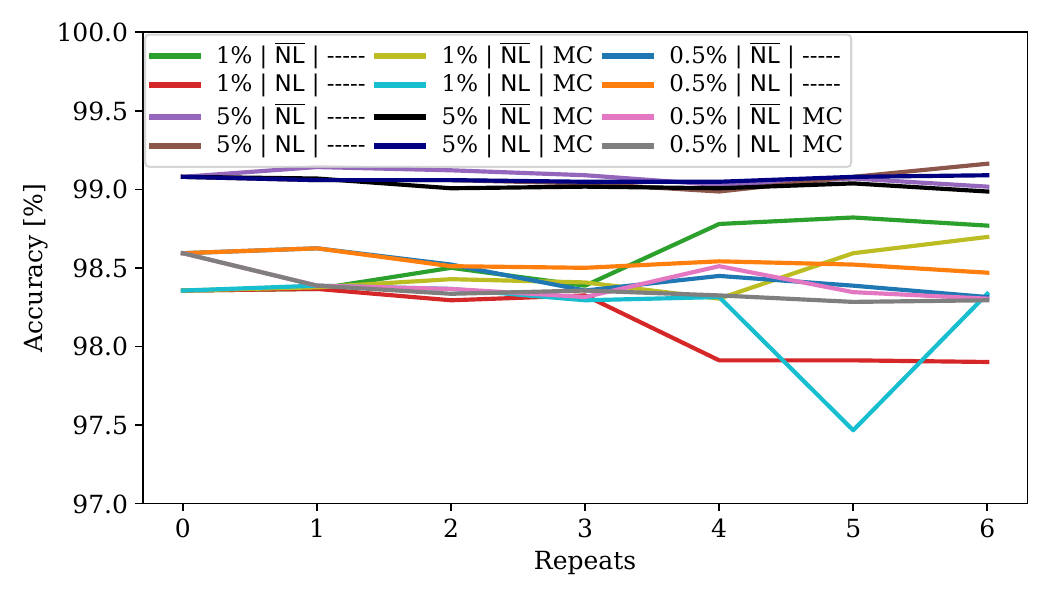}
        \end{minipage}
        \hfill
    	\begin{minipage}[t]{\y\linewidth}
            \centering
        	\includegraphics[trim=10 10 10 10, clip, width=1.0\linewidth]{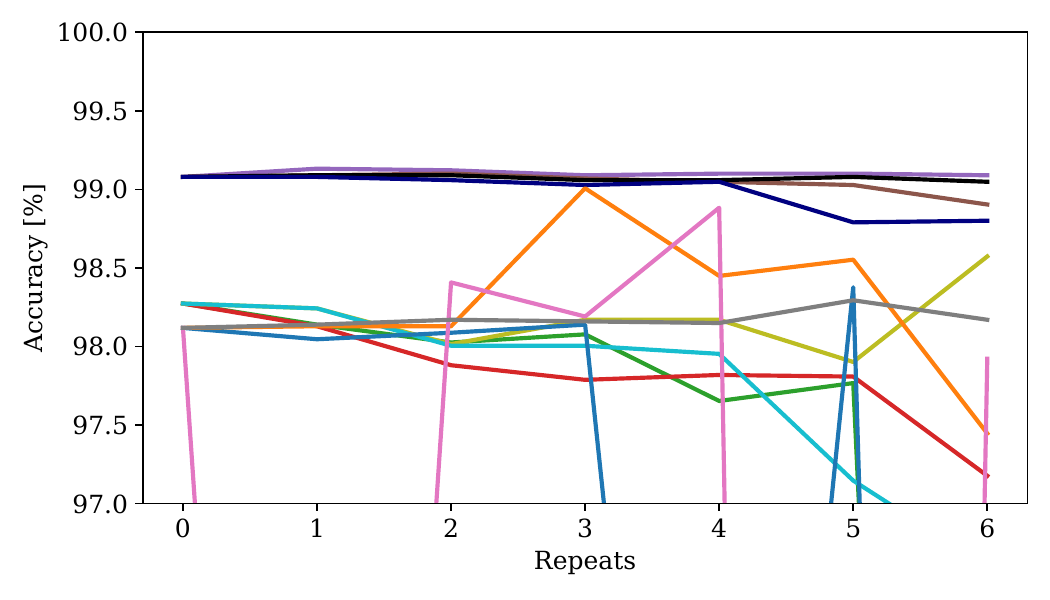}
        \end{minipage}
        \hfill
    	\begin{minipage}[t]{\y\linewidth}
            \centering
        	\includegraphics[trim=10 10 10 10, clip, width=1.0\linewidth]{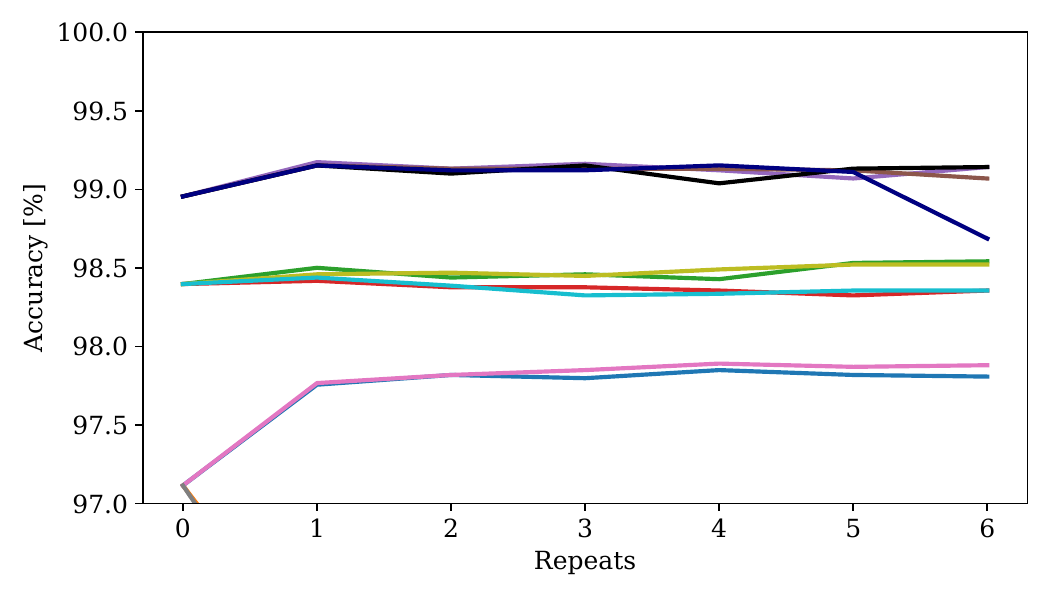}
        \end{minipage}
        \vspace{-0.2cm}
        \subcaption{Method proposed by Rizve et al.~\cite{rizve_duarte}.}
        \label{figure_evaluation1}
        \end{minipage}
	\begin{minipage}[t]{1.0\linewidth}
    	\begin{minipage}[t]{\y\linewidth}
            \centering
        	\includegraphics[trim=10 10 10 10, clip, width=1.0\linewidth]{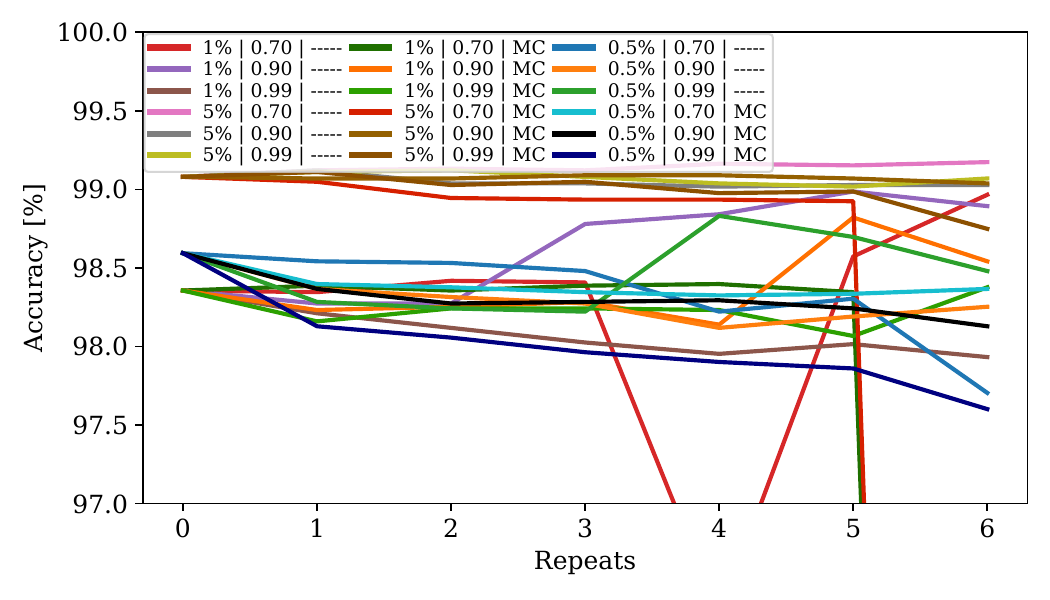}
        \end{minipage}
        \hfill
    	\begin{minipage}[t]{\y\linewidth}
            \centering
        	\includegraphics[trim=10 10 10 10, clip, width=1.0\linewidth]{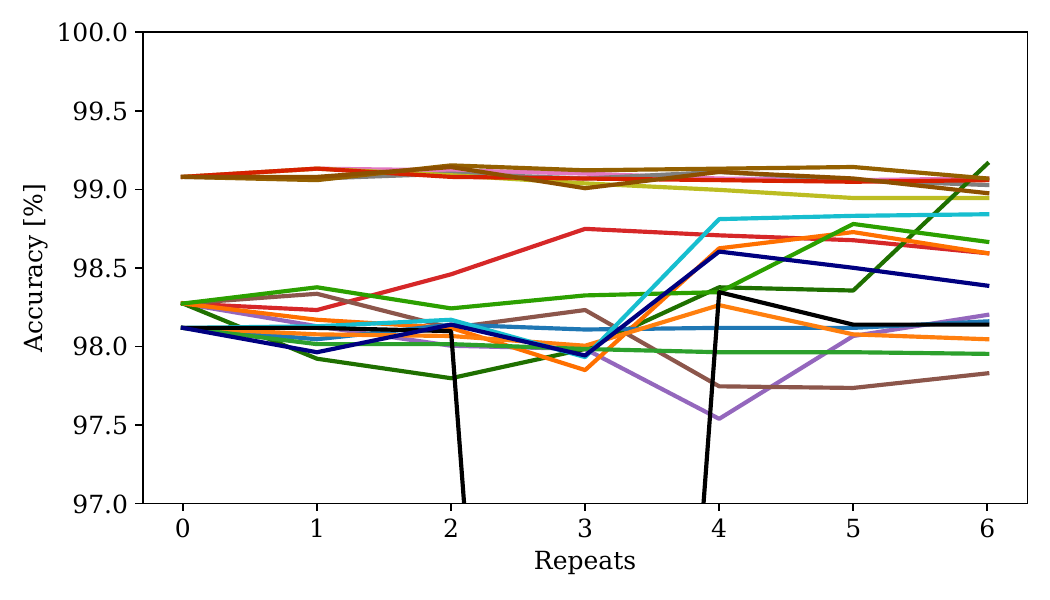}
        \end{minipage}
        \hfill
    	\begin{minipage}[t]{\y\linewidth}
            \centering
        	\includegraphics[trim=10 10 10 10, clip, width=1.0\linewidth]{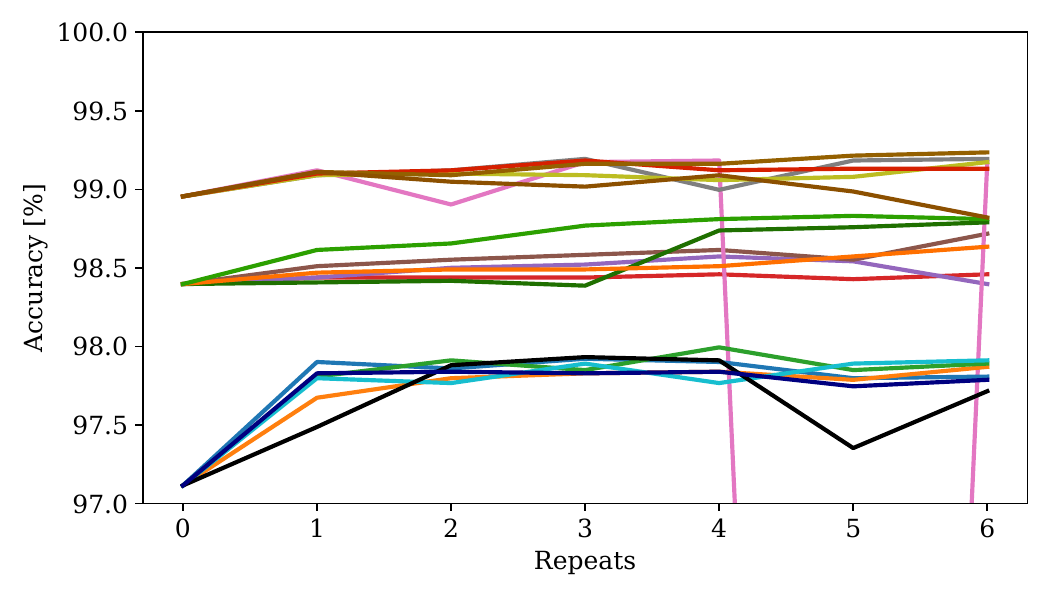}
        \end{minipage}
        \vspace{-0.2cm}
        \subcaption{Our pseudo-label selection based on softmax voting by four models.}
        \label{figure_evaluation2}
    \end{minipage}
    \vspace{-0.2cm}
    \caption{Evaluation results for the highway dataset 1 by pre-training on the highway dataset 2 (left), the controlled high-frequency dataset (middle), and the low-frequency dataset (right).}
    \label{figure_evaluation}
\end{figure*}

\paragraph{Evaluation Results.} In Figure~\ref{figure_evaluation}, we present the evaluation results for adaptation from three different datasets to the real-world highway dataset 1 over seven repetitions, both with and without negative learning (NL) and our voting mechanism based on softmax values. Four models were evaluated using varying proportions of labeled samples: 0.5\% (61 negative and 5 positive classes), 1\% (121 negative and 9 positive classes), and 5\% (605 negative and 41 positive classes). We also assess the effectiveness of pseudo-label selection using either MC techniques or a combination of MC and Deep Ensembles, with softmax thresholds $\gamma \in \{0.7, 0.9, 0.99\}$. The results demonstrate that without NL, using 1\% of labeled data yields significantly better performance (refer to Figure~\ref{figure_evaluation1}, left, green and red lines). However, with 5\% labeled data, NL becomes advantageous due to the higher volume of labeled samples, where noise has a more pronounced impact, and NL provides a mitigating effect. Furthermore, NL positively influences performance when combined with MC dropout. Regarding the dataset, with 5\% labeled data, the pre-training dataset becomes less relevant, as we consistently achieve 99.1\% accuracy. When using 0.5\% and 1\% labeled data, models exhibit greater robustness when adapting from the highway 2 dataset to highway 1, likely due to the real-world-to-real-world adaptation. In contrast, adaptation from controlled indoor environments to real-world scenarios demands more data. Three repetitions are generally sufficient for training a well-converged model due to the rapid convergence rate, although early stopping methods may be considered in future work. As expected, increasing the quantity of labeled data significantly improves accuracy, with larger labeled datasets (i.e., 5\%) leading to higher accuracy overall. However, even with just 0.5\% labeled data, we achieve accuracies ranging from 98.5\% to 99.0\%, comparable to the results obtained with 5\% labeled data. With respect to the softmax threshold, $\gamma = 0.9$ yields the best results. A threshold of $\gamma = 0.99$ selects too few samples, while $\gamma = 0.7$ selects too many and includes incorrect samples. We also evaluated the impact of using Deep Ensembles and MC dropout. The results indicate that using only Deep Ensembles is either superior or comparable to the combination of Deep Ensembles with MC dropout. Therefore, we recommend using only Deep Ensembles, depending on the dataset. In summary, our method outperforms the approach by Rizve et al.~\cite{rizve_duarte}, achieving a final accuracy of 99.2\%. Notably, even with only 0.5\% or 1\% labeled data, we obtain significantly higher accuracies. For 0.5\% labeled data, additional repetitions prove beneficial (see Figure~\ref{figure_evaluation2}, middle).

\begin{figure*}[!t]
    \centering
    \begin{minipage}[t]{0.325\linewidth}
        \centering
        \includegraphics[trim=10 10 10 10, clip, width=1.0\linewidth]{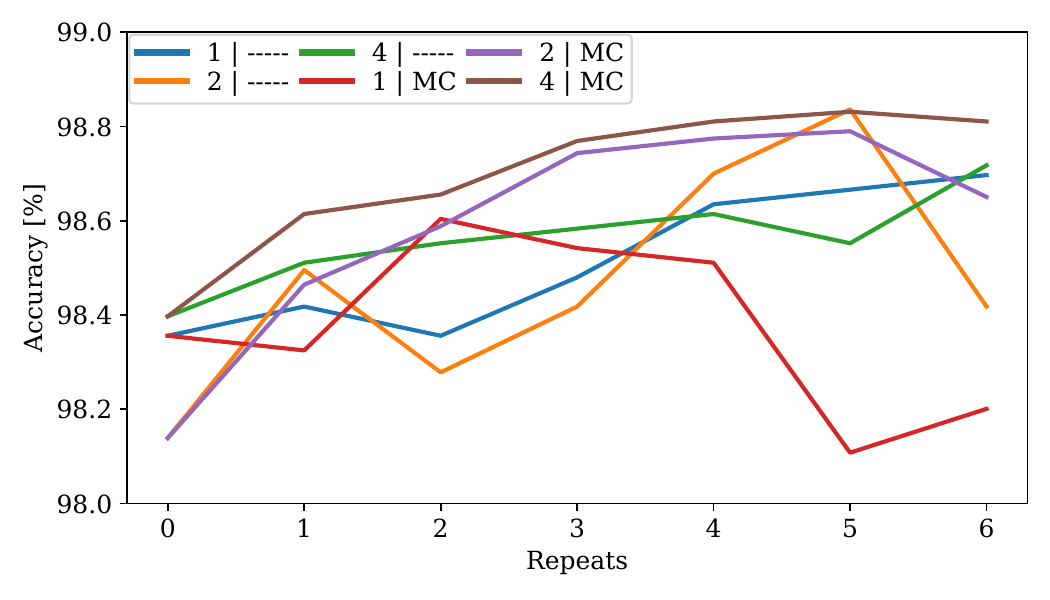}
        \vspace{-0.6cm}
        \caption{Evaluation of the number of models.}
        \label{figure_hyp1}
    \end{minipage}
    \hfill
    \begin{minipage}[t]{0.325\linewidth}
        \centering
        \includegraphics[trim=10 10 10 10, clip, width=1.0\linewidth]{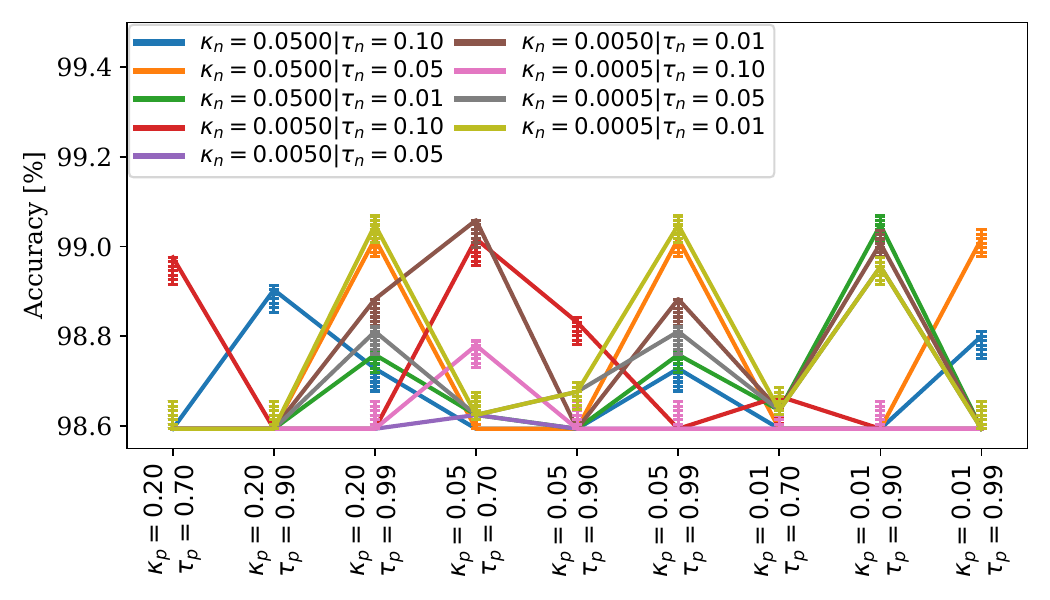}
        \vspace{-0.6cm}
        \caption{Evaluation of parameters $\kappa_p$, $\kappa_n$, $\tau_p$, and $\tau_n$.}
        \label{figure_hyp2}
    \end{minipage}
    \hfill
    \begin{minipage}[t]{0.325\linewidth}
        \centering
        \includegraphics[trim=8 10 10 10, clip, width=1.0\linewidth]{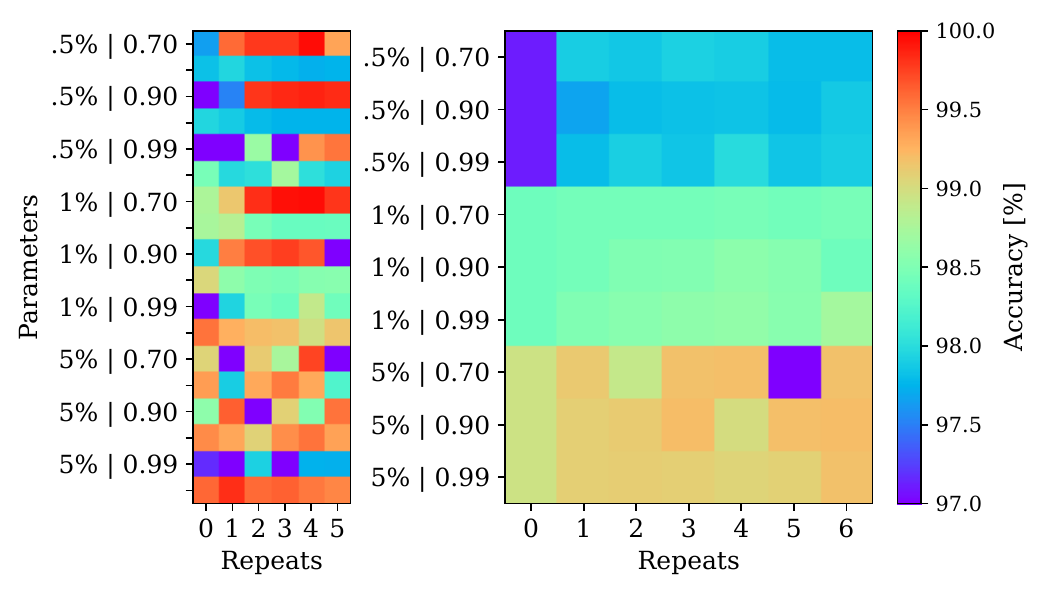}
        \vspace{-0.6cm}
        \caption{Number of samples selected for retraining.}
        \label{figure_hyp3}
    \end{minipage}
\end{figure*}

\paragraph{Evaluation of Number of Models.} In the following, we conduct hyperparameter searches for adaptation from the highway dataset 2 to the highway dataset 1. For comparison, refer to Figures~\ref{figure_hyp1}, \ref{figure_hyp2}, and \ref{figure_hyp3}, alongside Figure~\ref{figure_evaluation} (left). The importance of the number of models is evaluated in Figure~\ref{figure_hyp1}, both with and without MC dropout. The results indicate that using four models yields significantly better performance than using only one or two models. Notably, Deep Ensembles demonstrate greater robustness for voting with a single model compared to the combination of Deep Ensembles and MC dropout. As a result, we conduct the subsequent hyperparameter searches using four models.

\paragraph{Evaluation of Hyperparameters.} We present a hyperparameter search for the uncertainty thresholds $\kappa_p$ and $\kappa_n$, as well as the confidence thresholds $\tau_p$ and $\tau_n$, in Figure~\ref{figure_hyp2}. For the negative confidence threshold $\tau_n$, a value of 0.01 yields the best performance, while a low value of 0.0005 is optimal for the negative uncertainty threshold $\kappa_n$. For the positive thresholds, we select $\tau_p = 0.99$ and $\kappa_p = 0.05$ for subsequent training. Ultimately, we achieve an accuracy of 99.08\%, which is slightly lower than the performance of our proposed method shown in Figure~\ref{figure_evaluation2}.

\paragraph{Selected Samples.} We also evaluate the number of selected samples for each repetition in Figure~\ref{figure_hyp3}. The right plot presents the final classification accuracy in relation to the percentage of labeled data and the softmax threshold. Once again, it is evident that a larger amount of labeled data is crucial, with 5\% of labeled data yielding higher classification accuracy compared to 0.5\% and 1\%. Additionally, accuracy improves as the number of repetitions increases. A softmax threshold of $\gamma = 0.9$ remains the optimal choice. The left plot illustrates the percentage of samples where all four models predict the same class label, which are subsequently used for the pseudo-label dataset (upper row), as well as the percentage of these samples that are correctly classified (lower row). With a softmax threshold of 0.99, significantly fewer samples are selected for pseudo-labeling compared to thresholds of 0.7 and 0.9. Although the labeled samples selected at 0.99 are of higher quality, this does not result in an improved training process. A softmax threshold of 0.9 provides the best trade-off between the quantity of selected samples and the proportion of correctly labeled samples.
\section{Conclusion}
\label{label_conclusion}

In the context of GNSS interference classification within complex real-world environments, we proposed a semi-supervised pseudo-labeling methodology to facilitate adaptation from controlled indoor settings to real-world environments, as well as to accommodate changes in scenarios. Our approach integrates Deep Ensembles with MC dropout to mitigate the impact of noise encountered by the network and to produce an output distribution that remains invariant to perturbations and augmentations. We achieve an accuracy of 99.2\%, and our model demonstrates robustness even when utilizing only 0.5\% of labeled data (equivalent to 5 positive samples).

\section*{Acknowledgements}

This work has been carried out within the DARCII project, funding code 50NA2401, supported by the German Federal Ministry for Economic Affairs and Climate Action (BMWK), managed by the German Space Agency at DLR and assisted by the Bundesnetzagentur (BNetzA) and the Federal Agency for Cartography and Geodesy (BKG).

\bibliography{POSNAV2024}
\bibliographystyle{IEEEtran}

\end{document}